\documentstyle[aasms4,12pt]{article}
\raggedright
\newcommand{\mo}{$\rm \mu _{0}$}

\newcommand{\mss}{mag arcsec$^{-2}$}

\newcommand{\plm}{$\pm$ }

\newcommand{\lta}{$\leq $}
\newcommand{\gta}{$\geq $}

\newcommand{\Bmo}{\rm ${\mu _B}$(0) }

\newcommand{\muo}{\rm ${\mu}$(0) }

\newcommand{\alp}{$\alpha$\ }
\newcommand{\etal}{{\it et.al.}\ }

\newcommand{\kms}{km sec$^{-1}$\ }

\newcommand{\eg}{{\em e.\ g.\ }}
\newcommand{\ie}{{\em i.\ e.\ }}

\input{epsf}
\input{rotate}
\begin{document}
\title{\bf The Space Density of Galaxies through $\bf \mu_B(0)$ = 25.0 mag arcsec$^{-2}$}
\author{K. O'Neil}
\affil{Arecibo Observatory, HC03 Box 53995, Arecibo, PR 00612}
\affil{\it koneil@naic.edu}
\and
\author{G. Bothun}
\affil{Physics Dep't, University of Oregon, Eugene, OR 97402}
\affil{\it nuts@bigmoo.uoregon.edu}
\begin{abstract}
Using the catalog of O'Neil, Bothun, \& Schombert (1999),
we examine the central
surface brightness distribution ($\phi(\mu_B(0))$) of galaxies
in the 22.0 \lta \Bmo \lta 25.0 \mss\ range.  Taking advantage
of having a catalog in which each galaxy has a known central
surface brightness, scale length, and redshift, we apply
a bi-variate volume correction to the data and extend
the surface brightness distribution function 
by one magnitude, to 25.0 B \mss. The result 
is a flat (slope = 0) surface brightness distribution function
from the Freeman value of 21.65 \plm 0.30 to the survey limit of 25.0
\mss, more than 11$\sigma$ away.  This indicates that a significant
percentage of the baryonic content of the universe is likely in potentials
only dimly lit by the embedded galaxy.

\end{abstract}
\keywords{galaxies: luminosity function -- galaxies: statistics -- galaxies: low surface brightness}

\section{Introduction}

Our ability to detect objects in the Universe has been likened to 
standing in a well-lit room in the middle of the 
night, and trying to look through the window to describe the garden 100 yards away.
Although it may be possible to definitively say that the garden exists, and even
describe some of the large, well-defined plants, coming up with a quantitative
description of the fainter, or smaller, or more distant plants is an extremely 
difficult task.   Moreover, of primary scientific interest is not the garden
itself, but rather the evolutionary history of the plants that  occupy it.  
With only this one view of the garden available to us, it would be extremely
unlikely that our derived evolutionary history would be very accurate.  The
parallel between the garden and galaxy detection should be clear.  In
1965 Arp attempted to quantify this limited view of the universe by defining a 
``band of visibility,'' outside of which we are unable to
discern galaxies, due to either the galaxies small apparent size or optically diffuse
nature.   Arp's argument was later quantified by Disney (1976), showing that 
the visibility bias was rather severe.

Since Arp's work was published, we have successfully broadened the
``band of visibility'' through improvements in both instruments 
and detection techniques.  As an example, the superb angular resolution of 
HST has allowed for the distinction between true stars and galaxies which appear
star-like in lower resolution surveys (\eg Lilly, \etal 1995; Ellis, \etal 1996; Cowie, \etal 1996;
Steidel, \etal 1996;  Abraham, \etal 1996; Morris, \etal 1999; O'Neil, Bothun, \& Impey 1999).  The fact that most of these newly resolved galaxies are very far away
means that the local Universe is not filled up with little, dinky, high
surface brightness galaxies.
On the other end of the spectrum, both improvements in detection techniques (i.e.
Malin 1978; Schwartzenberg, \etal 1995) and the advent of CCD cameras as a tool in observing
has allowed for the detection of increasingly diffuse (low signal-to-noise) stellar systems
(\eg Impey, Bothun, \& Malin 1988; Davies, Phillipps, \& Disney 1988; Dalcanton,
\etal 1997; Schombert \& Bothun 1988; O'Neil, Bothun, \& Cornell 1997; Matthews \& Gallagher 1997).
Indeed the recent detection of extremely low surface brightness (LSB) dwarf
spheroidal galaxies around Andromeda by Armandroff, Davies, \& Jacoby (1998) is consistent
with the local Universe having a large population of low mass, nearly invisible
galaxies.   The Andromeda discovery underscores the severity of surface brightness
selection effects.  Where once the Milky Way stood alone in the Local Group as
a unique host of 7 LSB dwarf spheroidals, we now have detected an apparently
equivalent population around M31.

As new observational techniques broaden the visibility band, thus allowing new
objects to become detected, the issue shifts from an existence proof to
establishing the true space density of these newly discovered galaxies.
In establishing this, we must understand the survey limitations.  To compensate 
for these limitations, corrections must be made for the decreased probability
of detecting a galaxy the closer it lies to the survey limits.
The mathematical formalism of this correction has been both extensively discussed
in the literature over the last 20 years (i.e. Disney 1976; Disney \&
Phillipps 1983;  McGaugh 1996; de Jong 1996), and applied to the available data
(i.e. McGaugh, Bothun, \& Schombert 1995; de Jong 1996).  These preliminary
applications suggest that the space density of galaxies as a function of
their central surface brightness as measured in the blue (\Bmo) is relatively
flat or at most slow falling out to \Bmo = 23.5 B \mss.  However, Dalcanton
\etal 1997 present data which suggests that space density continues to rise
pass this limit.   We emphasize that these results refer exclusively to
{\it non-dwarf} galaxies; i.e. those objects with scale lengths larger than
$\sim$ 1 kpc.

In this paper we use the sample of LSB 
galaxies in the O'Neil, Bothun and Schombert (1999) catalog to determine the
surface brightness distribution function from a \Bmo of 22.0 B \mss\ through 25.0 B \mss\
(the catalog limits).  In addition to extending the known distribution by 
at least one \mss, the
O'Neil, Bothun, \& Schombert survey has the advantage of having
both well defined survey limits and known \Bmo, scale lengths, and redshifts,
allowing for the use of a bivariate correction for the survey selection
and the accurate determination of the surface brightness distribution function
in the 23.0 -- 25.0 B \mss\ range. The paper is laid out as follows:
Section 2 of describes the volumetric correction applied 
to the data, section 3 discusses the overall form of our determined surface 
brightness distribution, and section 4 briefly describes the implications of
these results.

\section{The Volume Correction}

As has often been discussed, detecting high surface brightness (HSB) galaxies
within a survey is considerably easier than detecting low
surface brightness (LSB) galaxies
(i.e. Freeman 1970; Disney 1976; Disney \& Phillipps 1983; McGaugh, Bothun,
\& Schombert 1995; Davies 1990; de Jong 1996; Bothun, Impey, \& McGaugh 1997;
Dalcanton \etal 1997).  Thus determining the true
(underlying) surface brightness distribution
of a sample of galaxies requires accounting 
for the probability of a galaxy being detected
by a survey of a given design.  For a field galaxy survey,
the probability of detection is determined simply by the available volume
which can be sampled for a galaxy of a given size and luminosity (e.g.
its surface brightness).  The volume corrected surface
brightness distribution is thus
\begin{equation} \phi(\mu_0)\:=\:\sum_{i=1}^{N}\:{S^i\over V^i_{max}}\end{equation}
where {\it i} is summed over all N galaxies in the sample, $S^i$ is 0 or 1 depending on
whether a galaxy lies within the described volume,
and $V_{max}={{4\pi}\over3}d_{max}^3$, the maximum volume in which a galaxy
could be detected.

For a surface brightness limited sample (i.e. the catalog of O'Neil, Bothun,
\& Schombert 1999, OBS from now on) $d_{max}$ can be found by 
requiring that the diameter
of the galaxy be equal to, or greater than, the minimum detectable diameter
($\theta \:=2r\:\geq\:\theta_l$).  
For a galaxy with an exponential surface brightness profile this gives:
\begin{equation} \mu(r)\:=\:\mu_0\:+\:1.086{r\over\alpha} \end{equation}
\begin{equation} \theta\:=\:2r\:=1.84\:\alpha\:(\mu_l\:-\:\mu_0)\:\propto\:
{h\over d}(\mu_l\:-\:\mu_0)\end{equation}
\begin{equation} d_{max}(\mu_0)\:\propto\:{h\over\theta_l}(\mu_l\:-\:\mu_0)\:.\end{equation}
where \muo is the central surface brightness of the galaxy, \alp is its scale
length in arcsecs, and h is its scale length in kpcs.  Thus, for a surface brightness
limited sample,
\begin{equation} V_{max}(\mu_0)\:\propto\:\left({h\over\theta_l}\right)^3(\mu_l\:-\:\mu_0)^3
\:\propto\:\left({{\alpha\:d}\over\theta_l}\right)^3(\mu_l\:-\:\mu_0)^3\:.
\label{eqn:volcor}\end{equation}

\section{The Surface Brightness Distribution, $\phi(\mu_0)$}

Figure~\ref{fig:lumfct} shows the results of applying the correction given in 
equation~\ref{eqn:volcor} to the O'Neil, Bothun and Cornell (1997) data using
the redshifts available in OBS.
The limiting diameter was set to 25'', and $\mu_{l}$ = 25.0 \mss.  This
corresponds to an approximate minimum physical diameter of 3 kpc (so again,
these are {\it non-dwarf} galaxies).  For the OBS survey,
the limiting central surface brightness was found through an extensive series
of computer models, in which Monte-Carlo-type simulations of the
images were created and searched for galaxies (O'Neil, Bothun, \& Cornell 1997).
As the true underlying galaxy
distribution of the computer-generated images were known, the detection cut-off
could be well determined.  As such, $\mu_{l}$, and thus V$_{max}$
for the OBS catalog is extremely well known.

Because the OBS sample is not uniformly distributed in space, but
instead follows the same large scale structure as the high surface
brightness (HSB) galaxies in the region (i.e. Figure 2 of OBS),
performing a V/V$_{max}$ test on the galaxies, and normalizing the
distribution function to that (i.e. de Jong 1996) would be extremely
difficult and possibly misleading at best.   In practice, the OBS sample
lies in a shell bounded by radial velocities of 4000 and 12000 \kms. 
The data for
this sample, as well as for the comparison samples, have therefore been normalized
to one (Figure~\ref{fig:lumfct}).  Additionally, to
insure against bias due to under sampling within a bin, the data from
OBS was binned to 0.5 \mss.  The errors bars for this data are simply $\sqrt N/N$.  The low
values for the surface brightness distribution between 22 -- 23
\mss\ are artificial, caused by the 22.0 B \mss\ cut-off in the survey sample
imposed in the OBS catalog.  This was not corrected for.

The data from this survey extends the faint end of the surface brightness
distribution function in a horizontal line from 23.0 \mss\ through 25.0 \mss,
the survey cut-off, matching the predictions made by, \ie\ McGaugh (1996) and 
Impey \& Bothun (1997).
Our distribution does, however, contradict
some of the data points of both de Jong (1996) and Davies (1990) in the 23.0 --
24.0 \mss\ range, where the de Jong and Davies samples appear to dip downwards.

The first, and most obvious, explanation for the discrepancy between our data
and that of both Davies (1990) and de Jong (1996) is that the volumetric corrections
for one or more of these samples was done incorrectly, either due to mis-identifying
the selection limits of the survey or through poor statistical sampling.
The OBS sample is designed to look for galaxies above the 22.0 \mss\ range, and is
therefore complete through 24.0 \mss, with the uncorrected data having a
flat surface brightness distribution from 22.5 through 24.0 \mss\ (\ie\
Figure 8 of O'Neil 1998) .  Additionally, the
surface brightness and diameter cut-off for the OBS sample was determined through
computer modeling (O'Neil, Bothun, \& Cornell 1997) and therefore is well determined.
In contrast, the de Jong sample ranges in central surface brightness from approximately
20.0 \mss\ through 24.1 \mss, with the the majority of the galaxies lying in the
21.0 -- 22.0 \mss\ range.  The volumetric corrections for the de Jong sample
are concerned only with the diameter limit ($\theta_{l}$) imposed on the
survey and do not account for potential surface brightness selection effects,
an omission which de Jong states could cause his survey to be undersampled at the
faint \muo, large scale length and/or small scale length, bright \muo ends of
the spectrum.  Combined with  the survey's under-sampling in the \muo $>$ 22.5 \mss\
range, this could result in an artificial drop in the de Jong surface brightness
distribution.

Like the de Jong (1996) sample, the Davies (1990) sample is concerned with the
entire range of central surface brightnesses, but in this case the total
number of galaxies involved in the survey should preclude any difficulties
with under-sampling.  The Davies sample was corrected for surface brightness
selection effects, with a limiting central surface brightness $\mu_{lim}(0)$ =
25.6 \mss, and $\theta_{l}$=7''.  This low value for $\theta_{l}$ potentially
mixed dwarfs and non-dwarfs together which could greatly confuse the
situation.  More importantly, no galaxies were actually detected near the
defined survey limits.  Thus it is entirely possible that 
the chosen sample limits simply do not accurately reflect the nature of
the survey and thus are inappropriate in determining the volumetric correction.
This could account for the apparent under-sampling in the 23.25 \mss\ bin
compared to our data.

Figure~\ref{fig:lumerr} show the
results of changing the binning for the OBS sample, from bins of 1.0 \mss\
through bins of 0.3 \mss.  The behavior of the surface brightness distribution
as the data becomes undersampled imitates the behavior of the de Jong
and Davies samples.  It is therefore possible that, as both the de Jong and 
Davies samples are primarily HSB galaxy samples, they are relatively undersampled
in the lower surface brightness regions. 

At this point the importance of the
chosen value for $\mu_{lim}$ should also be noted.  Choosing a value which
is fainter (or brighter) than the true survey limits will result in an artificial
lowering (raising) of the surface brightness distribution slope at faint \mo.
This is not surprising as it simply is a statement that if a survey is believed
to extend to, say, 26.0 \mss\ and yet detects no objects with \mo \gta 25.0
\mss, it would be accurate to assume that a fall-off in galaxy number density
at faint (\mo \gta 25.0) surface brightness is occurring.  The OBS sample 
is no exception to this rule.  Were $\mu_{lim}$ reduced to 26.0, a slight
decline in the slope of the surface brightness distribution function,
beginning at 24.0 \mss\ would be evident.   As  $\mu_{lim}$ was carefully determined for
the OBS sample, though, it should be an accurate representation of the survey's
true limitations.  With this, and the above, considerations in mind, it is likely
the flat surface brightness
distribution given by the OBS sample through 24.5 -- 25.0 \mss\ is an accurate
representation of the surface brightness distribution in the local (z$<$0.05)
universe.   The implication of such a flat distribution remains profound.

\section{LSB Galaxies and the Baryon Fraction}

In 1991 Walker, \etal used the latest nuclear cross sections
to calculate the baryon abundance (D, $^3$He, $^4$He,
and $^7$Li) within the framework of the standard hot big bang
cosmological model.  Their calculations led to a nucleon-photon
ratio of
\[2.8\:\leq\:n_b/n_\gamma\:\times\:10^{10}\:\leq\:4.0\]
and a baryon density parameter of $\Omega_B\:h^2_{100}$ = 0.0125 \plm 0.0025.
Estimating the known baryon mass density of the universe using
$\Omega_B$ = $\Omega_{E/S0}$ + $\Omega_{Sp}$ + $\Omega_{clusters}$ +
$\Omega_{groups},$  Persic and Salucci (1992)
found $\Omega_B$ = 2.2 + 0.6$h^{-3/2}_{100}$ x 10$^{-3}$, showing that
70\% -- 80\% of the predicted baryon mass does not even exist
in standard galaxy catalogs.  

In their review, Impey \& Bothun (1997) speculate that
if the M/L of LSB galaxies is higher than in HSB galaxies,
as is indicated in most LSB galaxy studies (i.e. McGaugh \& de Blok 1997;
OBS),
the faint end of the luminosity function has a slope of $-$1.6 -- $-$1.8
(i.e. de Propris, \etal 1995; Bothun, Impey, \& Malin 1991), and 
the central surface brightness distribution remains flat
through 28.0 -- 30.0 B \mss, then the total contribution to the 
baryonic mass from galaxies is $\Omega_B\:h^2_{100}$ = 0.014 -- 0.020,
well within the bounds set by Walker, \etal.
Whether the underlying distribution is flat through 28.0 B \mss\ or 
falls off after  26.0 B \mss\ awaits deeper data to determine.  Nevertheless,
the results of this study fortify those of McGaugh, Bothun, \& Schombert (1995) and McGaugh
(1996) and greatly suggest that a lot of baryons are contained in potentials
that host very diffuse and hard to detect galaxies.

\section{Conclusion}
Using a bivariate volume correction, we extend the surface brightness
distribution function in a horizontal line from the Freeman value of
21.65 B \mss\ through 25.0 B \mss, the limit of the OBS catalog.
This result is consistent with previous studies (e.g. McGaugh, Bothun, \& Schombert 1995;
Dalcanton \etal 1997) but extends them to fainter surface brightness levels.
Our result is somewhat inconsistent with the findings of two previous surveys
in the 23.0 -- 24.0 B \mss\ range (Davies 1990; de Jong 1996).  However,
our survey was designed specifically to detect galaxies in this surface
brightness range and we have quite well defined survey limits
($\theta_l,\:mu_l$).  This leads us to have considerable confidence in our results.
It can therefore be said with confidence that LSB galaxies are 
both important in their own right, and are very significant
contributors to the total baryonic mass in the universe.

\clearpage
\centerline{REFERENCES}

Abraham, R..G. Valdes, F., Yee, H., van den Bergh, A.  1994, ApJ, 432, 75\\

Armandroff, T., Davies, J., Jacoby, G. 1998 AJ 110, 2287\\

Arp, H. 1965 ApJ 142, 402\\

Bothun, G., Impey, C., \& McGaugh, S. 1997 PASP 109, 745\\

Bothun, G., Impey, C., \& Malin, D. 1991 ApJ 376, 404\\

Cowie, Lennox L., Songaila, Antoinette, Hu, Ester M., \& Cohen, J. G. 1996 AJ 112, 839\\

Dalcanton, J., \etal 1997  AJ 114, 635\\

Davies, J. 1990 MNRAS 244, 8\\

Davies, J., Phillipps, S., Disney, M. 1988 MNRAS 231, 69\\

de Jong, R. 1996 A\&A 313, 46\\

de Propris, R., Pritchet, C., Harris, W., \& McClure, R. 1995 ApJ 450, 524\\

Disney, M., \& Phillipps, S. 1983 MNRAS 205, 1253\\

Disney, M. 1976, Nature 263, 573\\

Ellis, Richard S., Colless, Matthew, Broadhurst, Tom, Heyl, Jeremy, \&
Glazebrook, Karl 1996 MNRAS, 280, 235\\

Freeman, K. 1970 ApJ 160, 811\\

Impey, C., \& Bothun, G. 1997 ARA\&A 35, 267\\

Impey, C., Bothun, G., \& Malin, D. 1988 ApJ 330, 634\\

Lilly, S. J., Le Fevre, O., Crampton, David, Hammer, F., \& Tresse, L. 1995 ApJ 455, 50\\

Malin, D. 1978 Nature 276, 591\\

Matthews, L., \& Gallagher, J. 1997 AJ 114, 1899\\

McGaugh, S., \& de Blok, W.J.G. 1997 MNRAS 290m 533\\

McGaugh, S. 1996  MNRAS 280, 337\\

McGaugh, S., Bothun, G., \& Schombert, J. 1995 AJ 110, 573\\

Morris, S. \etal 1999 ApJ, in press\\

O'Neil, K., Bothun, G., \& Impey, C. 1999, preprint\\

O'Neil, K., Bothun, G., \& Schombert, J. 1999, preprint\\

O'Neil, K. 1998 Ph.D. Dissertation University of Oregon:Eugene\\

O'Neil, K., Bothun, G., \& Cornell, M 1997 AJ 113, 1212\\

Persic, M. \& Salucci, P. 1992 MNRAS 258, 14\\

Schombert, J., \& Bothun, G. 1988 AJ 95, 1389\\

Schwartzenberg, J., \etal 1995 MNRAS 275, 121\\

Steidel, C., Giavalisco, M., Pettini, M., Dickinson, M., Adelberger, K, 1996 ApJ 462, 17\\

Walker, T., \etal 1991 ApJ 376, 51\\

\clearpage

\centerline{FIGURES}

\figcaption[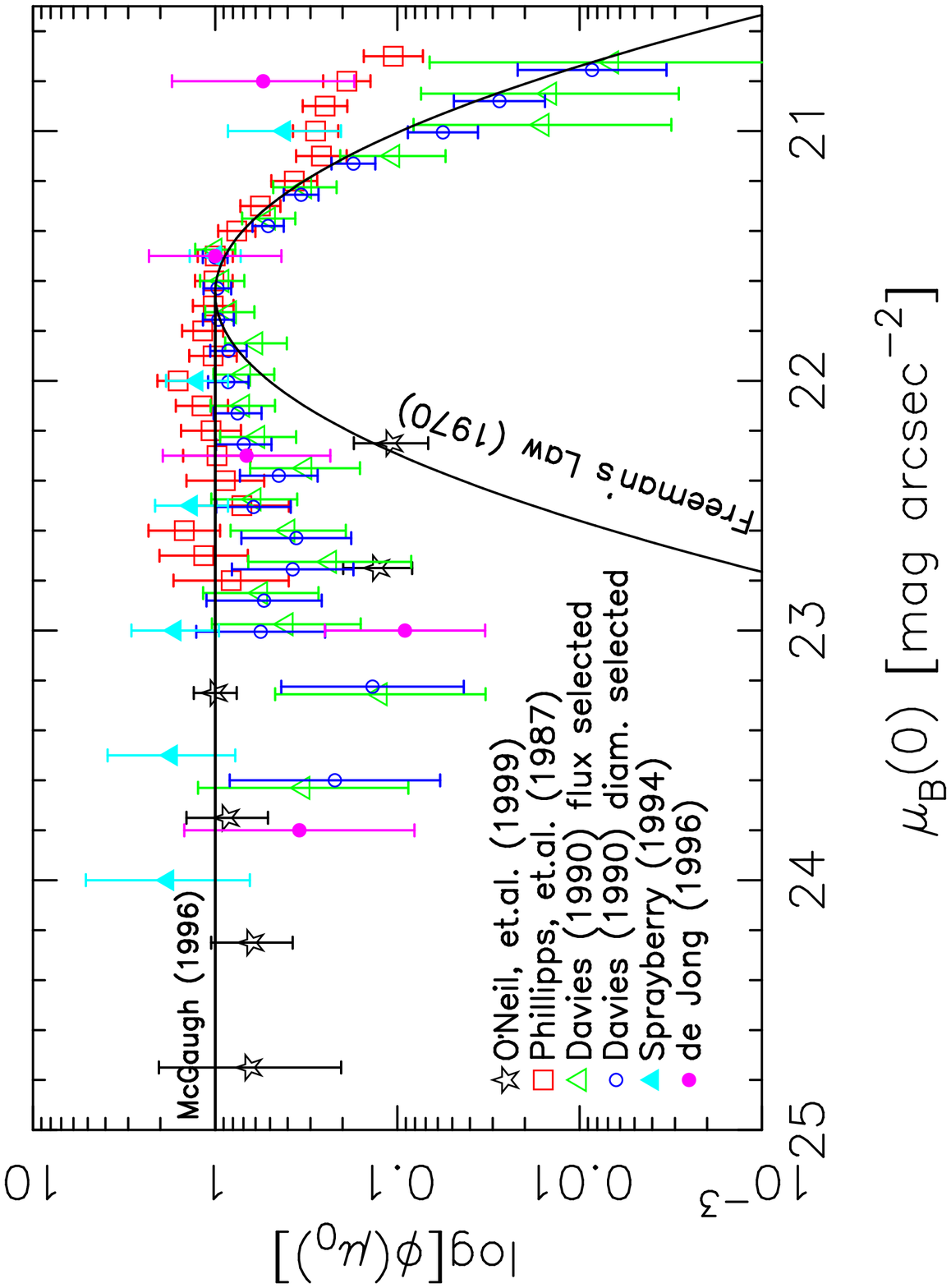] {The volume corrected surface brightness distribution
function for all the galaxies in this and other surveys.  \label{fig:lumfct}}

\figcaption[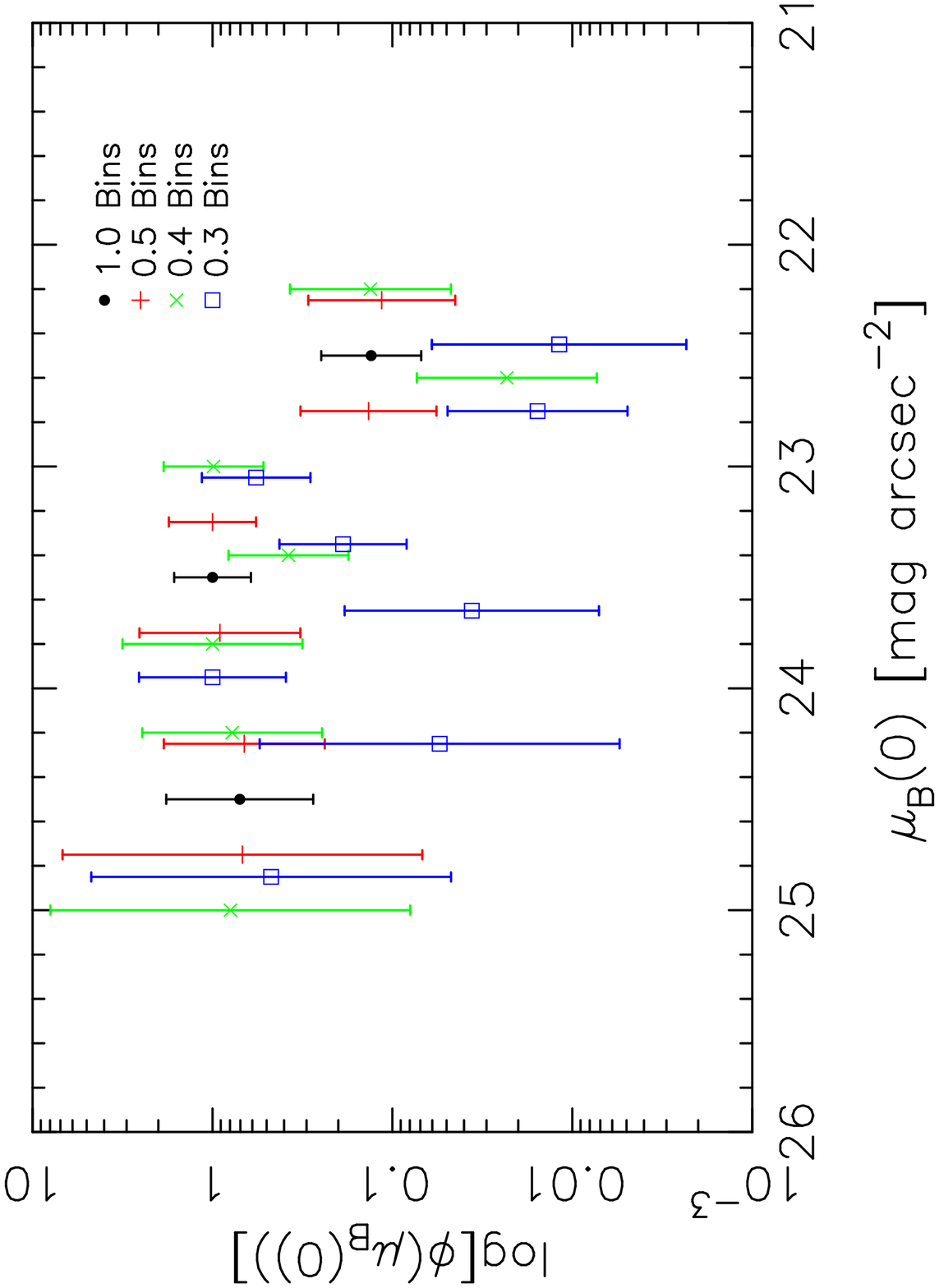] {The effects of under-sampling the data in this survey.
The circles, crosses, Xs, and squares result from binning the galaxies in
1.0, 0.5, 0.4, \& 0.3 \mss\ bins, respectively. \label{fig:lumerr}}

\end{document}